\documentclass[fleqn,10pt]{wlscirep}
\usepackage[utf8]{inputenc}
\usepackage[T1]{fontenc}
\usepackage{color,soul}
\usepackage{mathptmx}
\usepackage{float}
\usepackage{amsmath}
\usepackage{amsmath}
\usepackage{blindtext}
\usepackage{gensymb}
\usepackage{booktabs,chemformula}
\usepackage{sidecap}
\usepackage{verbatim}
\usepackage{graphicx}
\usepackage{tablefootnote}
\usepackage{footnote}

\usepackage{dcolumn}
\usepackage{bm}
\usepackage{times}
\usepackage{subfigure}
\usepackage{lineno}
\usepackage{hyperref}
\usepackage{color}
\usepackage{babel}

\title{Toward accurate thermal modeling of phase change material based photonic devices}

\author[1,*]{Kiumars Aryana}
\author[1]{Hyun Jung Kim}
\author[2]{Cosmin-Constantin Popescu}
\author[3]{Steven Vitale}
\author[4]{Hyung Bin Bae}
\author[4]{Taewoo Lee}
\author[2,5]{Tian Gu}
\author[2,5]{Juejun Hu}
\affil[1]{NASA Langley Research Center, Hampton, 23666 VA, USA}
\affil[2]{Department of Materials \& Science Engineering, Massachusetts Institute of Technology, Cambridge, MA, USA}
\affil[3]{Lincoln Laboratory, Massachusetts Institute of Technology, Lexington, MA, USA}
\affil[4]{KAIST Analysis Center, Korea Advanced Institute of Science and Technology, Yuseong-gu, Daejeon 34141, Korea}
\affil[5]{Materials Research Laboratory, Massachusetts Institute of Technology, Cambridge, MA, USA}

\affil[*]{kiumars.aryana@nasa.gov}

\begin{abstract}
Reconfigurable or programmable photonic devices are rapidly growing and have become an integral part of many optical systems. The ability to selectively modulate electromagnetic waves through electrical stimuli is crucial in the advancement of a variety of applications from data communication and computing devices to environmental science and space explorations. Chalcogenide-based phase change materials (PCMs) are one of the most promising material candidates for reconfigurable photonics due to their large optical contrast between their different solid-state structural phases. Although significant efforts have been devoted to accurate simulation of PCM-based devices, in this paper, we highlight three important aspects which have often evaded prior models yet having significant impacts on the thermal and phase transition behavior of these devices: the enthalpy of fusion, the heat capacity change upon glass transition, as well as the thermal conductivity of liquid-phase PCMs. We further investigated the important topic of switching energy scaling in PCM devices, which also helps explain why the three above-mentioned effects have long been overlooked in electronic PCM memories but only become important in photonics. Our findings offer insight to facilitate accurate modeling of PCM-based photonic devices and can inform the development of more efficient reconfigurable optics.
\end{abstract}

\begin{document}
\flushbottom
\maketitle

\renewcommand{\thefootnote}{\roman{footnote}}

\section{Introduction}
Reconfigurable optics ranging from zoom lenses to tunable optical filters have recently garnered great interests for more compact and energy efficient systems \cite{julian2020reversible, shalaginov2021reconfigurable, chaudhary2019polariton, kim2023p}. One of the quintessential examples of the necessity for reconfigurable optics lies in the most powerful telescope ever built, \textit{The James Webb Space Telescope (JWST)}. At the heart of JWST there is a near-IR camera that enables imaging a wide variety of electromagnetic spectra from 0.6 to 5 $\mu$m. In order to selectively choose different electromagnetic wave spectra, there are 29 different passive optical filters sitting on two separate filter wheels which are mechanically swapped depending on the operation mode \cite{gardner2006james, beichman2012science, beichman2014observations}. As amazing as this technology is, it requires bulky and complex components with mechanically moving parts that are not ideal for applications in space. Ideally, for adjusting the operation mode and imaging at different wavelengths, all 29 filters should be replaced by one unique tunable filter that can be electrically switched between different passband wavelengths with high speed and fidelity \cite{gu2022reconfigurable}. Light detection and ranging (LiDAR) is another technology that could benefit from non-mechanical light modulators for higher efficiency beam steering and scanning capabilities \cite{kim2021heterogeneously}. Thus far, depending on the application, a wide variety of techniques have been proposed to modulate the electromagnetic waves using thermo-optical \cite{harris2014efficient, bosch2019polarization}, electro-optical \cite{xu2005micrometre}, magneto-optical \cite{tzuang2014non, takagi2014magneto}, opto-mechanical \cite{arbabi2018mems}, and acousto-optical effects \cite{sun2016optical}. Nonetheless, there is still a dire need for more efficient non-mechanical tunable optics that are fast, robust, and compatible with standard semiconductor foundry fabrication processes. 


One of the potential material candidates for reconfigurable optics is chalcogenide-based phase change materials (PCMs) that undergo a solid-state phase transformation upon thermal stimuli. These materials have large optical property contrast between their amorphous and crystalline states, which makes them suitable for modulating electromagnetic waves. Another important feature of PCMs is their analog nature which enables a continuous range of properties depending on the amorphous to crystalline ratio in the material. For instance, it has been demonstrated that GST can be used as a coating material to thermally camouflage objects from the background via its variable emissivity at different phases \cite{qu2018thermal}. More recently, PCMs have been integrated into metasurfaces in order to develop tunable properties \cite{wang2016optically, julian2020reversible, yang2022reconfigurable, wang2023varifocal, ding2019dynamic, de2020tunable, abdollahramezani2022electrically}.

To better understand the operation of these devices and rationally guide their design, considerable efforts have been dedicated to precise modeling of PCMs' phase transition behavior\cite{scoggin2018modeling, faraclas2014modeling}. However, to date, the vast majority of the investigations have focused on electronic memory configurations, which do not apply to photonic devices where electrothermal switching via external micro-heaters rather than direct current injection is employed\cite{zhang2021myths}.

In addition, for the application of PCMs in functional photonic devices, it is preferred to switch a larger volume of PCM to increase its overlap with the optical mode, thereby maximizing the optical contrast \cite{abdollahramezani2020tunable, fang2021non}. While this is less of a concern for memory applications where the volume of phase change units are on the order of a few nanometers \cite{kim2010high, kang2014considerations}, achieving reversible switching of PCM at large length scales on the order of microns or greater presents several challenges in comparison to conventional nanoscale memory cells. These challenges include slower heat dissipation, greater atomic migration, higher power consumption, and potential delamination, all of which can negatively impact the uniformity and durability of the phase transition\cite{aryana2021interface, popescu2023learning}. Therefore, thermal management plays a critical role in improving the performance of PCM-based devices, especially as their length scales increase.

Furthermore, while several studies have explored thermal and kinetic modeling of the PCM crystallization process\cite{redaelli2008threshold, sebastian2014crystal, carrillo2021system, wang2021scheme,meyer2020multiphysics}, the amorphization transition has received much less attention. We argue that the ability to accurately characterize the amorphization process is equally--if not more--important. Amorphization involves raising the PCM above its melting point ($>$600\degree C) followed by rapid cooling ($\sim$10$^{9}$–10$^{12}$ K s$^{-1}$) \cite{salinga2018monatomic, salinga2013measurement}, which is a far more thermally vigorous process than crystallization. Therefore, the reliability and endurance of PCM devices are largely impacted by amorphization\cite{martin2022endurance}. It has also been shown that the amorphization process can influence the crystallization speed in the subsequent cycle \cite{orava2012characterization, schumacher2016structural,salinga2013measurement}, and thus it plays a consequential role in the repeatability of the switching process, especially during multi-level operation\cite{li2019fast, zhang2019miniature}.

In this paper, we examined the impact of three previously largely overlooked effects--the enthalpy of fusion, temperature-dependent heat capacity, as well as the thermal conductivity of liquid-phase PCMs--on its phase transformation. The results presented here would enable accurate modeling of PCM devices while providing insight into the development of large-scale, robust, and energy-efficient PCM-based devices.

\begin{figure}[htb]
\begin{center}
\includegraphics[scale=0.50]{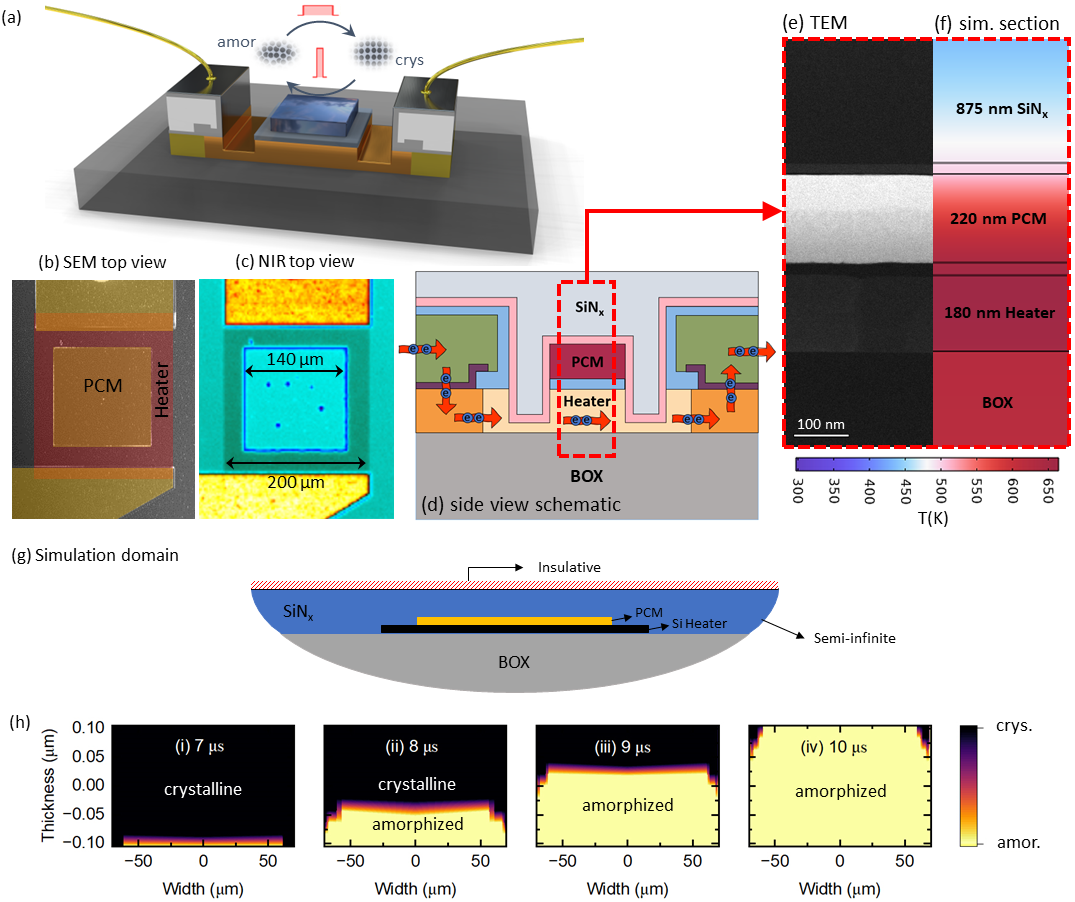}
\caption{This figure illustrates the configuration of PCM-based device studied in this paper. Panel (a) provides a 3D schematic of the device configuration, while panels (b) and (c) show top view images obtained using scanning electron microscopy (SEM) and near-infrared (NIR) camera (FLIR A6262) with a wavelength range of 600 - 1700 nm. Panels (d-f) depict cross-sectional view of the device with all the layers adjacent to the PCM, accompanied by corresponding transmission electron microscopy (TEM) image and simulation sections that are the focus of this study. The simulation domain, along with the relevant boundary conditions used in this study, are presented in panel (g). (h) The expansion of the amorphous region within the crystalline state due to thermal excitation from the heater underneath. The time stamp denotes the duration since the initiation of the heating pulse. Note that the images are not drawn to scale.
\label{fig:1a}}
\end{center}
\end{figure} 

\section{Device Configuration and Thermal Modeling}
In our models, we have adopted a doped Si micro-heater design similar to that reported by Rios \emph{et al.} \cite{rios2022ultra}. Figure \ref{fig:1a}(a) illustrates 3D schematic of the heater device configuration and the layer structures. The top-view image of the fabricated device is presented in Fig. \ref{fig:1a}(b,c) using two imaging techniques, scanning electron microscopy (SEM) and near-IR (NIR) camera. In this configuration, the PCM pixel size is 70\% of the underneath Si heater, with lateral sizes of 140 $\mu m$ and 200 $\mu m$, respectively. We choose Si heaters because they are transparent in both the near-infrared (NIR) and mid-infrared (MIR) regimes and have shown to have reliable stability \cite{zhang2019miniature, zheng2020nonvolatile, rios2022ultra}. This enables operation in the transmissive mode instead of reflective, a critical factor in the development of reconfigurable optics such as meta-lenses, filters, and beam steering. Moreover, Si heaters are also compatible with standard foundry manufacturing \cite{vitale2022phase} to facilitate scalable deployment of PCM-based devices. In order to switch PCM between amorphous and crystalline phases, we use a function generator and a power supply to deliver the high current necessary to heat up the Si heater to sufficiently high temperatures. The resistance of the devices varies in the range of 40-50 $\Omega$, which requires voltages from 36 V to 44 V for amorphization depending on the device resistance. The conclusions drawn from our study are equally applicable to other electrothermal PCM switching configurations as well.

\setlength{\tabcolsep}{.7em}
\begin{table}[htb]
\caption{\label{tab:01} Room temperature properties of materials used in the simulations.}
  \centering
  \begin{tabular}{@{}cccccc@{}}
    \toprule
    \toprule
    Film & Thickness & Thermal Conductivity & Specific Heat & Density \\
         &     (nm)      & (W m$^{-1}$ K$^{-1}$) & (J kg$^{-1}$ K$^{-1}$) & (kg m$^{-3}$ \\
    \midrule
    SiN$_x$  & 875 nm  & 2.0\cite{braun2021hydrogen}     & 1631   & 3100    \\
    Al$_2$O$_3$ & 25 nm   & 2.0\cite{cappella2013high}     & 880   & 3950    \\
    GSST   & 210 nm &  $\kappa(T)$ \cite{aryana2021suppressed}    & 240\cite{zhao2022exploring, aryana2021suppressed}   & 6200    \\
    SiO$_2$ & 30 nm   & 1.35\cite{regner2013broadband}    & 1000   & 2650    \\    
    Si heater & 180 nm   & 135\cite{regner2013broadband}     & 710   & 2329    \\

    \bottomrule
    \bottomrule
  \end{tabular}
\end{table}

In order to examine the thermal transport and temperature distribution within the proposed device architecture, we utilized finite element simulation software (COMSOL Multiphysics). A 2D model of the device was created based on its dimensions, which were established through scanning transmission electron microscopy and are illustrated in Fig. \ref{fig:1a}(b,e). The device's cross-section schematic was used as the basis for the model, as shown in Fig. \ref{fig:1a}(d). Our primary focus in this study is investigating the amorphization process, so we assumed that the initial state of the PCM is fully crystalline, unless stated otherwise. The simulations were performed at room temperature, with constant temperature set to 24 $\degree$ C at the boundaries of the simulation domain far away from the device (Fig.\ref{fig:1a}(g)). To ensure accurate modeling, we selected a domain size 10 times larger than the device width, and we validated this assumption by testing matrix lengths ranging from 10 to 50 times larger, finding no impact on the device's temperature distribution. Given the microscale dimensions and fast thermal transport timescale in addition to the exceptionally low thermal conductivity of air compared to other materials within the device, it is a reasonable assumption to disregard convection and radiation effects and set the top surface as an insulative boundary condition. Further, we assumed a constant thermal boundary conductance of 100 MW m$^{-2}$ K$^{-1}$ across all interfaces to capture interfacial thermal resistance between different materials \cite{aryana2021interface}. The thermal conductivity of the PCM layer, denoted as $\kappa(T)$, was assumed to vary with temperature based on Refs. \cite{aryana2021suppressed}, while the remaining material thermal properties for the simulated non-PCM layers were considered to be temperature-independent.

In our investigation of the amorphization process, we take Ge$_2$Sb$_2$Se$_4$Te (GSST) which is a widely used PCM for optical applications due to its low loss absorption in the infrared regime\cite{zhang2017broadband,zhang2019broadband}. We presume that once the temperature of the PCM surpasses the melting point 600 \degree C, the liquid part completely transitions into the amorphous phase once cooled down. The simulation results depicted in Fig. \ref{fig:1a}(h) shows the progression of the amorphized region in relation to the base crystalline state upon thermal excitation. It is important to note that the contour plots are not drawn to scale, as the width of the PCM is 1000 times greater than its thickness.

In order to experimentally illustrate the amorphization process, we deposited 140 $\mu m$ $\times$ 140 $\mu m$ of GSST on a 200 $\mu m$ $\times$ 200 $\mu m$ transparent silicon heater. To prevent oxidation and improve heat dissipation during the amorphization process, we encapsulated the device with 875 nm sputtered SiN$_x$. Using a short-wavelength IR camera (FLIR A6262), which operates in 600-1700 nm wavelength range, we monitored changes in the PCM's reflectivity as we sent successive amorphization pulses. The GSST is initially in the crystalline phase. Upon sending incremental amorphization pulses that are separated by 30 seconds, we ensured the device reached equilibrium at room temperature before each image was collected.

\begin{figure}[htb]
\begin{center}
\includegraphics[scale=0.50]{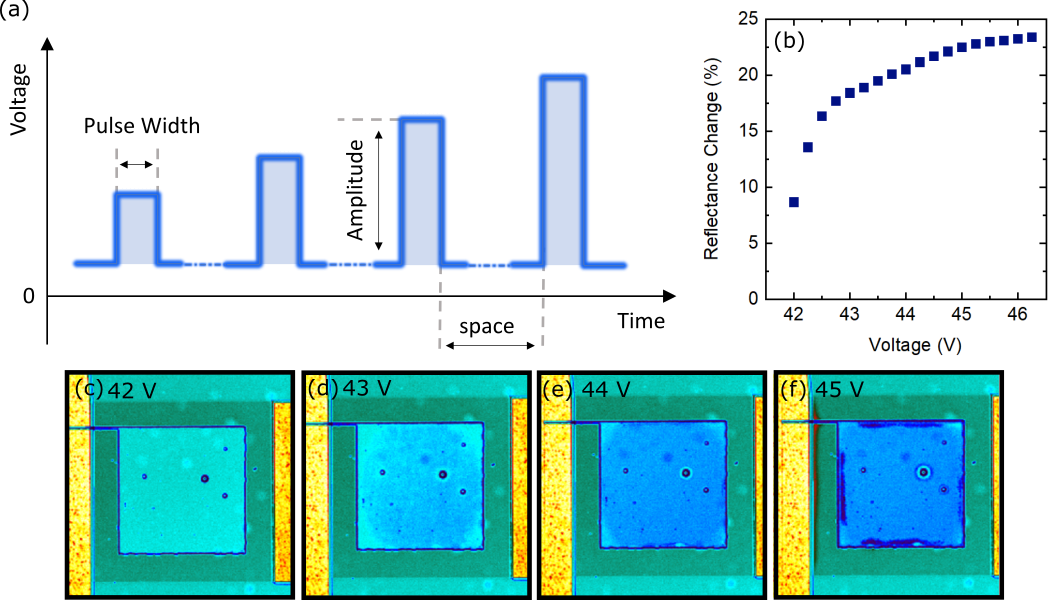}
\caption{(a) Incremental sequential pulses that are applied to GSST on a heater with 10 $\mu$s pulse width and 30 seconds interval period. (b) Changes in the reflectivity of the GSST with respect to base crystalline phase as a function of applied voltage.  (c-f) Images from the surface of the GSST upon application of different voltages and progression of the amorphous phase. 
\label{fig:4}}
\end{center}
\end{figure} 


Figure \ref{fig:4} shows the incremental pulses applied to the PCM and the percent change in the reflectivity of the surface with respect to the applied voltage. As can be seen in the images presented in Fig. \ref{fig:4}(c), the sample did not show any change for pulses with amplitudes up to 42 V. As soon as the voltage exceeded this threshold, the reflectivity of the center of the sample began to decrease which is indicative of partial amorphization. On the other hand, at the boundaries of PCM and near the corners, Fig. \ref{fig:4}(d), we observed no change in the color of PCM, which we believe is due to non-uniform temperature distribution across the PCM. This is consistent with our simulated temperature distribution in the PCM for low amplitude pulses that lead to partial amorphization, consistent with the results presented in Fig. \ref{fig:1a}(h). Upon increasing the pulse voltage, the change in the color became more pronounced and the unswitched crystalline regions closer to the edges began to amorphize. From these results we can see the reflectivity plateaus at 45.5 V where upon increasing the voltage no observable change in the reflectivity of the PCM was detected. By approximating the changes in the reflectivity of the PCM with respect to its initial crystalline phase, we observed an $\sim$23\% change in the reflectivity upon complete transformation to the amorphous state. In the following section, we will elaborate on several important factors that should be properly considered to enable accurate modeling of the progressive amorphization process shown here.

\section{Enthalpy of Fusion and Heat Capacity of Supercooled Liquids}

\begin{figure}[htb]
\begin{center}
\includegraphics[width=\columnwidth]{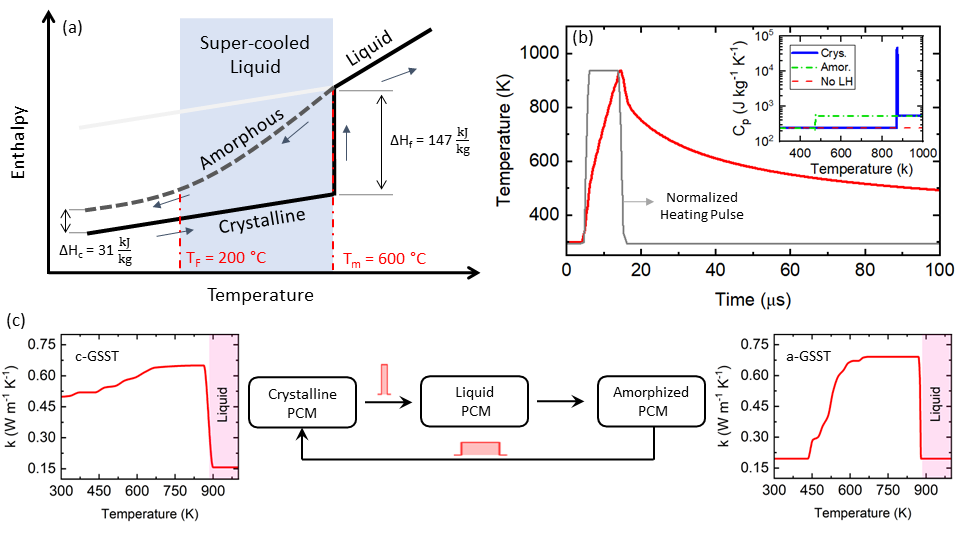}
\caption{(a) The enthalpy-temperature curves for GST, showcasing classical patterns observed during solid-liquid phase transition as well as glass transition, and (b) maximum temperature profile and pulse shape within the PCM during the amorphization cycle. The inset shows heat capacity trend as a function of temperature for crystalline and amorphous phases, as well as when latent heat (LH) of melting is not considered. (c) The switching cycle and the corresponding thermal conductivity trend as a function of temperature for the amorphous and crystalline phase of GSST. 
\label{fig:1b}}
\end{center}
\end{figure}

Figure \ref{fig:1b}(a) plots the classical enthalpy-temperature curves of a glass-forming solid. $\Delta H_c$ and $\Delta H_f$ denote the enthalpies of crystallization and fusion, respectively. The former gives the enthalpy difference between a solid's amorphous and crystalline states whereas the latter, also known as the latent heat of melting, represents the energy required to change the state of a crystalline solid unit mass to liquid at the melting point, without a temperature increase. The other important feature evidenced by this figure is that while the heat capacity of the crystal solid (represented by the slope of the curve) is relatively insensitive to temperature up to its melting point, the heat capacity of the amorphous phase increases considerably from the glassy state near room temperature to the supercooled liquid state past glass transition. This is attributed to the increasing degrees of freedom that the atoms can access in the liquid phase versus in the solid phase\cite{varshneya2013fundamentals}.

The significance of the enthalpy-temperature relation in the PCM switching process is two-fold. First, during amorphization, additional heat must be supplied to convert the solid phase into liquid and during the process, the temperature at the solid-liquid interface is held constant. This effect, even though not accounted for in some prior studies\cite{koelmans2015projected, rios2018controlled}, can significantly alter the temperature distribution and solid-liquid interface location. For PCMs, the $\Delta H_f$ ranges from 98 to 147 kJ/kg \cite{senkader2004models, zhao2022exploring, johnson1981enthalpies, pashinkin2008heat}, and the corresponding latent heat is equivalent to the energy needed to raise the temperatures by nearly 600 K. Therefore, the enthalpy of fusion is an important factor in thermal modeling especially when the PCM device size is large, as we will explain later.

The second implication is more subtle and to our knowledge has not been explicitly discussed in literature. In previous thermal simulations of PCM-based photonic devices\cite{zhang2019nonvolatile, zheng2020modeling,abed2020tunable,zhang2021comparison, zhang2021electrically, rios2021multi, lawson2022time, zhang2023nonvolatile}, the heat capacity of amorphous PCM has been quoted as a constant up to the melting point. However, as can be seen from Fig. \ref{fig:1b}(a), the heat capacity of amorphous PCM is not a constant and must increase when transitioning into the supercooled liquid state, or when heated to above its Fictive temperature. Assuming a constant heat capacity will underestimate the amount of energy needed to bring the amorphous PCM to the melting point, roughly by $\Delta H_f - \Delta H_c$. In PCMs, $\Delta H_c$ ranges between 31--65 kJ/kg\cite{zhao2022exploring, pustkova2006non, vcernovskova2014thermal, kalb2007calorimetric, tae2013comparison}. For GST, $\Delta H_f - \Delta H_c = $ 116 kJ/kg. The temperature-dependent heat capacity can therefore significantly impact thermal simulation results in PCM devices, as we will show later.

\begin{figure}[htb]
\begin{center}
\includegraphics[scale=0.4]{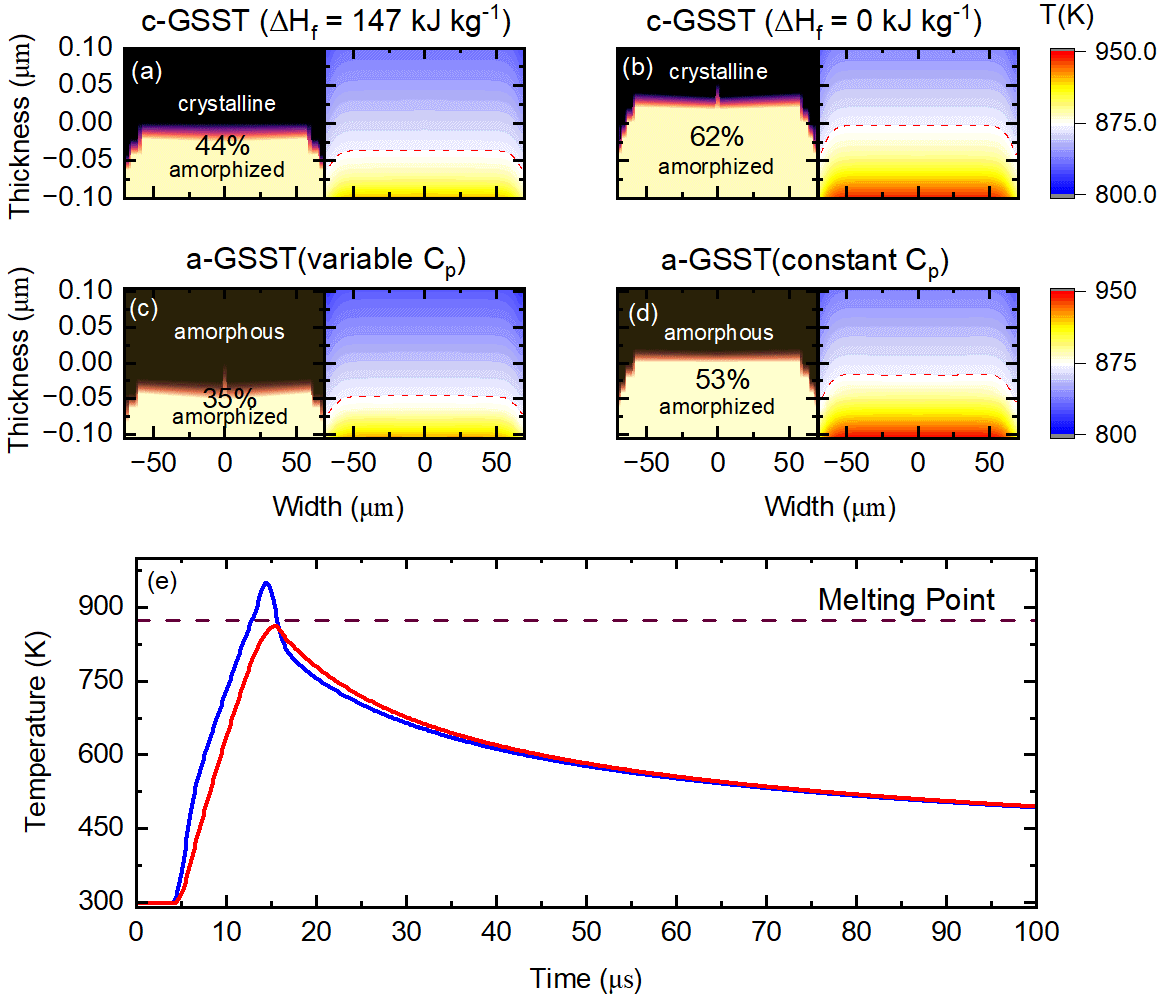}
\caption{This figure illustrates the phase transformation and temperature distribution in PCM for GSST after amorphization pulse under various conditions, including: (a) considering the enthalpy of fusion ($\Delta H_f$ = 147 kJ kg$^{-1}$), (b) disregarding the enthalpy of fusion ($\Delta H_f$ = 0 kJ kg$^{-1}$), (c) considering the enthalpy of crystallization ($\Delta H_f$ = 65 kJ kg$^{-1}$), and (d) the amorphous phase without considering the enthalpy of fusion ($\Delta H_f$ = 0 kJ kg$^{-1}$). The contour plots are not drawn to scale, as the width of the PCM is 1000 times greater than its thickness. The melting temperature threshold is denoted by the dashed red line in the 2D temperature contour. Additionally, the maximum and minimum temperatures in PCM and the corresponding temperature gradient with respect to time are shown in (e).
\label{fig:2}}
\end{center}
\end{figure} 

In order to demonstrate the impact of enthalpy of fusion on the phase transformation during the amorphization process, we consider two cases: one where latent heat of melting is taken into account ($\Delta H_f$ = 147 kJ kg$^{-1}$), which appears as a spike in heat capacity between 870 K and 876 K as shown in the inset in Fig. \ref{fig:1b}(b), and one where it is (erroneously) not considered ($\Delta H_f$ = 0). Figure \ref{fig:1b}(c) illustrates the switching cycles in PCM-based devices that undergo a liquid phase during amorphization, as well as the corresponding thermal conductivity trend as a function of temperature for both the crystalline and amorphous phases. 

Figures \ref{fig:2}(a-b) show the different modeling outcomes of the amorphization cycle in the GSST device in the two cases. The contour plots provide a comparison of the percentage of amorphized volume and temperature distribution for GSST when considering the latent heat versus ignoring it. When considering the effect of latent heat, Figs. \ref{fig:2} (a,b) illustrate that at the same input power, a smaller PCM volume undergoes solid-liquid phase transformation. This is because some of the energy in the crystalline phase is utilized to overcome the latent heat of melting, resulting in a lower temperature increase, less molten GSST, and consequently, less amorphization. Figures \ref{fig:2}(a,b) demonstrate that neglecting the impact of latent heat would result in a nearly 18\% overestimation of the amorphized area and temperature distribution in the PCM for an input power of 130 mW.

Next we examine the influence of the temperature-dependent heat capacity of PCMs. Unfortunately, experimental data on the heat capacity data of the (supercooled) liquid phase of PCMs are not available, and therefore we will need to make some educated assumptions. We start by assuming that the heat capacity remains the same as the room temperature value up to the Fictive temperature $T_F$, beyond which the heat capacity abruptly increases to the (supercooled) liquid phase value. This is a valid approximation as glass transition typically occurs over a narrow temperature window compared to the entire amorphization process. We further assume a Fictive temperature of $T_F$ = 200 \degree C, which corresponds to the approximate temperature range of glass transition measured in GST. If we further quote the $\Delta H_f$ = 147 kJ/kg and $\Delta H_c$ = 31 kJ/kg values from GST, then the heat capacity jump at $T_F$ can be obtained from $\Delta C_P$ = $(\Delta H_f - \Delta H_c)/(T_m - T_F)$, where $T_m$ represents the melting point. Following this, we take a constant heat capacity of 240 J kg$^{-1}$ K$^{-1}$ and 528 J kg$^{-1}$ K$^{-1}$ for below and above $T_F$, respectively. In this case, since we are studying the melting of an amorphous phase, we assume that the amorphization pulse is send to an already amorphous PCM. This may seem redundant, but it provides critical insights for investigating the thermal properties of the amorphous phase in multi-level programming, where a mixture of amorphous and crystalline phases is present. As shown in Fig. \ref{fig:2}(c-d), assuming a constant heat capacity as a function of temperature, similar to the crystalline phase, over-predicts the amorphization and temperature rise in the PCM. Figure \ref{fig:2}(e) shows the temperature rise as a function of time during crystalline-to-amorphous phase transition for location points in the GSST; center and closest to the heater (blue), and center farthest from the heater (red). According to this plot, the temperature gradient along the thickness of the GSST can reach as high as $\sim$104 K at the peak temperature. This large temperature gradient could lead to non-uniformity and reduced lifetime. Further, it is important to note that, in all cases, the crystalline phase leads to a higher amorphization rate in the PCM than the amorphous phase, which is attributed to the higher thermal conductivity of the crystalline phase compared to the amorphous phase. The subsequent section will discuss the importance of the thermal conductivity of PCM during the amorphization process.

\begin{figure}[htb]
\begin{center}
\includegraphics[scale = 0.4]{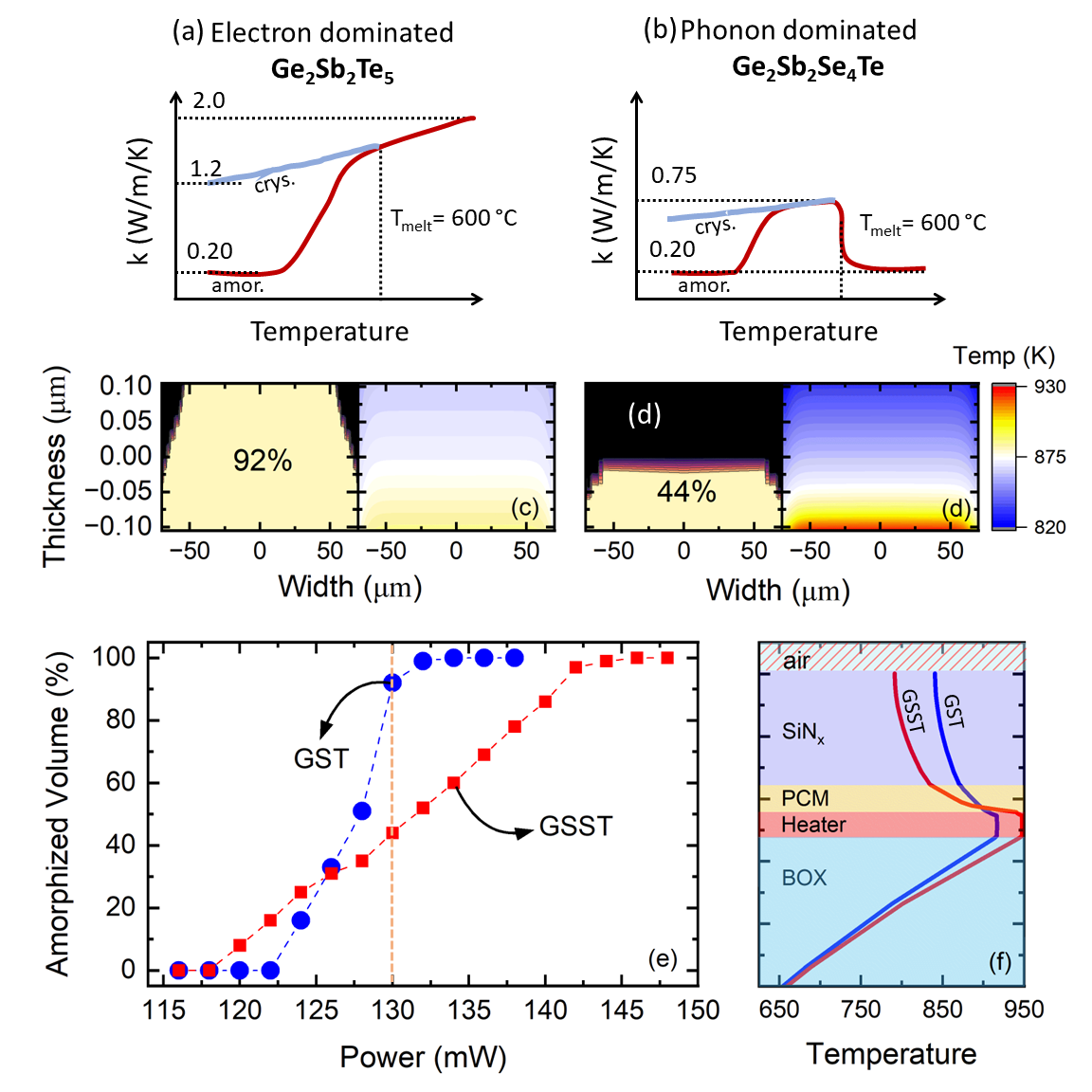}
\caption{(a,b) Qualitative thermal conductivity trend for GST and GSST with respect to temperature. The actual values for temperature dependant thermal conductivity that are used in the simulations can be found in Ref. \cite{aryana2021suppressed}. (c,d) PCM thermal conductivity effect on the phase transformation and temperature distribution. (e) Percentage of amorphized PCM at different input power for GST and GSST. (f) The through-plane temperature profile at the midpoint of the device for GSST and GST at the end of the amorphization pulse (after 10 microseconds).
\label{fig:3}}
\end{center}
\end{figure} 

\section{Thermal Conductivity of Liquid Phase}
From previous section, we observed that thermal conductivity of the PCM phase plays an important role in the degree of amorphization. For developing reliable devices with an extended lifetime, a uniform temperature distribution across the PCM during the amorphization cycle is a necessity. Theoretically, for triggering amorphization of the PCM, we need to deliver enough power to uniformly raise its temperature a few degrees above the melting point. Nonetheless, considering most PCMs have typically low thermal conductivity ($<$1 W/m/K), as the device length scale increases, the formation of a non-uniform temperature profile is inevitable. Using the right combination of materials and device design, we can mitigate this non-uniformity and potentially help improving the device durability. For the PCM with embedded heater configuration which is the focus of this study, we demonstrated that the center of the device at the interface between the PCM and heater reaches the highest temperature during the amorphization cycle leading to a flat-dome shaped amorphized region. Thus, for increasing the amorphized volume, typically the heater must operate tens of degrees higher than the melting temperature of the PCM to amorphize regions away from the heater. This, however, could be a major issue as a higher temperature would create a localized hot spot and impose a larger thermal strain on the device. In this section, we discuss the importance of PCM intrinsic thermal conductivity for a more uniform temperature distribution and greater phase transformation.

Although most phase change materials have intrinsically low thermal conductivity, depending on their compositions, they behave significantly differently at high temperatures, especially in the liquid phase. For instance, it has been shown that the thermal conductivity of fully crystalline GST is driven by electrons whereas in fully crystalline GSST, the electronic contribution is significantly suppressed and the thermal transport is driven by phonons \cite{aryana2021suppressed}. This means that the GST thermal conductivity increases with temperature as depicted in Fig. \ref{fig:3}(a) even in the liquid phase due to the increased contribution of electronic carriers which is also consistent with prior electrical conductivity measurements in the liquid phase \cite{endo2010electric, siegrist2011disorder, crespi2014electrical, baratella2022first}. On the other hand, the thermal conductivity of GSST is expected to drop to its amorphous value due to disruption of periodicity and emergence of disorder in the atomic structure (see Fig. \ref{fig:3}(b)). It is noted that there are no reports for the thermal conductivity of GSST in the liquid phase to validate or refute our hypothesis, yet, we can make an educated speculation about how GSST behave in the liquid phase. Considering this and using temperature-dependent thermal conductivities from Ref. \cite{aryana2021suppressed}, an estimation for the amorphization volume and temperature distribution in GST and GSST is presented in Fig. \ref{fig:3}(c-d). From these results, during the amorphization period, we observe a more uniform temperature distribution in the GST layer where the temperature gradient in the entire layer only reaches $\Delta$T = 31 K, whereas for the case of GSST it reaches $\Delta$T = 104 K. As a result of the large temperature gradient in GSST, heat transport and hence amorphization across its volume are hampered. This is evident from Fig. \ref{fig:3}(e), where the amorphization fraction in GSST consistently lags behind that in GST even for identical heater configuration and voltage pulse parameters, with an amorphization fraction difference up to 48\% at 130 mW power. Furthermore, Fig. \ref{fig:3}(f) displays the highest temperature attained at the end of the amorphization pulse (after 10 microseconds) for GSST and GST across various layers in the stack. As per the plot, the maximum temperature recorded for the GSST case is 947 K, whereas for GST, it is 916 K. Thus, if GST is employed as the PCM under identical conditions, the Si heater temperature would remain 31 K lower than in the case of GSST. This marked difference highlights the critical importance of quantifying liquid-phase thermal conductivity to allow for material-specific thermal management and reliable PCM switching in photonic devices.

\section{Switching Energy and Size Scaling in PCM-based Devices}
This section aims to examine the effect of scaling PCM-based devices on their energy consumption and overall performance. According to Fourier's law, the energy required to heat a material is proportional to its cross-sectional area (A), with Q = - kA(dT/dx), assuming a constant thermal conductivity (k) and temperature gradient (dT). Thus, doubling the area of a material results in doubling of the energy required to generate the same temperature gradient. Nonetheless, in a device configuration, it is not only the PCM that heats up during the switching cycles but also the surrounding materials. The rate of heat loss from PCM to the surrounding materials depends on its surface area, which increases less rapidly than the volume as the material grows in size. Consequently, smaller volumes like PCM-based memory cells in storage devices have a larger surface-to-volume ratio and more contact areas with the surroundings, resulting in a higher proportion of heat being transferred to the surroundings and more intense thermal leakage. This means that, the scaling of PCM-based devices is expected to follow a sub-linear trend. As the length scale of the PCM device increases (more quantitatively as the heater size reaches a few tens of microns), the switching energy is increasingly being dominated by the intrinsic properties of the PCM rather than those of surrounding materials. It therefore comes as no surprise that the aforementioned effects resulting from enthalpy of fusion, supercooled liquid heat capacity and liquid phase thermal conductivity have long been neglected in electronic memories and only become relevant in photonic devices. In order to investigate this, we use our finite element model and change the lateral dimension of the heater and PCM in order to assess the impact of size on the power consumption. 

\begin{figure}[htb]
\begin{center}
\includegraphics[scale=0.55]{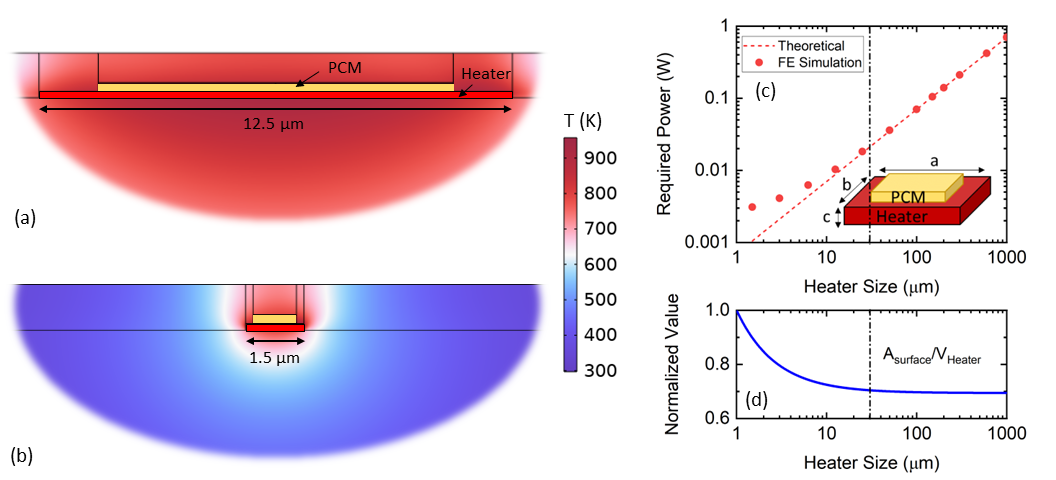}
\caption{Temperature distribution in the PCM device with embedded heater with lateral dimension of (a) 1.5 $\mu m$ and (b)12.5 $\mu m$. (c) The theoretical value and the simulation results for device power consumption with respect to the heater size from 1 to 500 $\mu m$. (d) Normalized ratio for surface area relative to volume as the lateral width of the Si heater increases from 1 to 1000 $\mu$m. In the FE simulation, the input power is adjusted to produce a similar temperature rise across all heater sizes.
\label{fig:44}}
\end{center}
\end{figure} 

Figures \ref{fig:44}(a,b) depict the temperature distribution during the amorphization pulse for two different device sizes. In this particular device architecture, the power consumption is determined by the heater size. To perform the simulations, the device size, or more accurately, the heater size, is varied from 1.5 $\mu m$, where the thickness of the PCM ($\sim$210 nm) is comparable to the device lateral dimension, to 300 $\mu m$, where the thickness of PCM is negligible compared to the heater lateral dimension. To enable a meaningful comparison of power consumption among devices with varying heater sizes, we select a 200 $\mu m$ heater as a benchmark. For this, an input power of 140 mW is required for a 200 $\mu m$ heater to attain 86\% amorphization and a maximum temperature rise of around 1000 K when subjected to a 10 $\mu s$ pulse. Upon changing the device size, we adjust the input power to get the same level of amorphization and temperature rise in the PCM. Theoretically, since Q and A have a linear relation, as the areal size of the heater decreases, we would expect a linear reduction in the power consumption, where for a heater with a smaller size of 1.5 $\mu m$, 1 mW of power is expected to generate the same temperature rise as in the 200 $\mu m$ heater. This is shown as the dashed line in Fig. \ref{fig:44}(c). However, simulations indicate that three times more power input (3 mW) is necessary for the 1.5 $\mu m$ heater to generate the same temperature rise and phase transformation in the smaller heater. In other words, our simulation results for various device lateral dimensions depicted by solid red circles reveal that the power consumption deviates from the linear trend as the heater size decreases below 30 $\mu m$. This deviation occurs due to the fact that the surface area to volume ratio is higher for smaller devices as depicted in Fig. \ref{fig:44}(d), resulting in greater thermal leakage from the boundaries to the surrounding environment. Consequently, creating the same temperature rise for the smaller heater size requires more power. This significant difference in power requirements is attributed to the larger surface area of smaller heaters relative to their volume, which leads to greater thermal leakage and, thus, a higher power demand to attain the same temperature rise. On the other hand, our results indicate that for heaters larger than 30 $\mu m$, the surface area to volume ratio remains relatively constant. Consequently, the power consumption follows a linear trend, as expected.

\hfill \break

\section{Conclusion}
In this study, we examined the amorphization process in large-area PCM-based photonic devices. The study specifically examined the impact of enthalpy of fusion, temperature-dependent heat capacity, thermal conductivity of liquid PCM, and scaling of device size on the phase transformation and power consumption. Our results demonstrated the importance of accounting for the enthalpy of fusion and the heat capacity change during glass transition when modeling the amorphization process. Furthermore, according to our simulations, in order to increase the dimensions of PCM-based photonic devices, PCMs with electron-dominated thermal conductivity such as Ge$_2$Sb$_2$Te$_5$ are more favorable as their thermal conductivity increases with temperature even after the solid-liquid phase transition. We compared the phase transformation degree and temperature distribution in the PCM layer for an electron-dominated PCM such as Ge$_2$Sb$_2$Te$_5$ and a phonon-dominated PCM such as Ge$_2$Sb$_2$Se$_4$Te and showed that the electron-dominated PCM leads to less temperature rise and more uniform temperature distribution in the PCM layer while providing more than two times higher amorphization volume. Furthermore, we discovered that for devices with lateral dimensions smaller than 30 $\mu$m, the power consumption does not scale linearly with the size of the device due to the increased ratio of PCM surface area to volume. However, for devices with lateral sizes larger than 30 $\mu$m, power consumption increases linearly with device size. Overall, our findings contribute to understanding and accurate modeling of thermal transport phenomena in PCM-based photonic devices and could lead to the development of reconfigurable metasurfaces with functional properties like filters, lenses, and beam steering.





\section{Methodology}
\noindent
\textbf{Transmission Electron Microscopy.} The specimen for STEM observation was prepared by lift out via ion-beam technology by using a Focused Ion-beam (FIB) system (Helios G5 UX, ThermoFisher Scientific). Protective amorphous carbon and Pt layers were applied over the region of interest before ion milling in the FIB system. HAADF STEM images were acquired with a transmission electron microscope (Titan cubed G2 60-300, ThermoFisher Scientific) at 300kV with a spherical aberration(Cs) corrector (CEOS GmbH). 

\noindent
\textbf{Device Fabrication.} The specimen for TEM and NIR optical analysis was fabricated on a doped Si-on-insulator platform. The GSST was deposited via thermal evaporation from a pre-weighted fresh source in a custom built deposition chamber. The film was subsequently encapsulated with 20 nm of $Al_2O_3$ via atomic layer deposition and further encapsulated with 875 nm of reactively sputtered silicon nitride.


\section{Disclaimer}
Specific vendor and manufacturer names are explicitly mentioned only to accurately describe the test hardware. The use of vendor and manufacturer names does not imply an endorsement by the U.S. Government nor does it imply that the specified equipment is the best available.

\bibliography{Refs.bib}

\section*{Data availability.} 
    The data that support the findings of this study are available from the corresponding author upon reasonable request.

\section*{Acknowledgements}
    This work was supported in part by the NSF Award Number 2132929. Research was sponsored by the National Aeronautics and Space Administration (NASA) through a contract with ORAU. The views and conclusions contained in this document are those of the authors and should not be interpreted as representing the official policies, either expressed or implied, of the National Aeronautics and Space Administration (NASA) or the U.S. Government. The U.S. Government is authorized to reproduce and distribute reprints for Government purposes notwithstanding any copyright notation herein.

\section*{Author contributions}
    K.A., H.J.K., and J.H. designed the study. K.A. performed the simulations. C.C.P, T.G., and J.H. fabricated the device. K.A. performed the electrical biasing measurement. H.B.B. and T.L. performed the SEM and TEM measurements. K.A. and J.H. wrote the manuscript.
    
\section*{Competing interests}   
    The authors declare no competing interests.
    
\noindent \textbf{Correspondence} and requests for materials should be addressed to K.A.

\end{document}